\documentstyle[12pt]{article}

\newcommand{\beq}{\begin{equation}}
\newcommand{\beqn}{\begin{displaymath}}
\newcommand{\zen}{\end{equation}}
\newcommand{\zenn}{\end{displaymath}}
\def\pnot{\mbox{${\not{\hbox{\kern-3.0pt$p$}}}$}}
\def\qnot{\mbox{${\not{\hbox{\kern-2.0pt$q$}}}$}}
\def\enot{\mbox{${\not{\hbox{\kern-2.0pt$e$}}}$}}
\def\knot{\mbox{${\not{\hbox{\kern-2.0pt$k$}}}$}}
\def\rnot{\mbox{${\not{\hbox{\kern-2.0pt$r$}}}$}}
\def\r'not{\mbox{${\not{\hbox{\kern-2.0pt$r'$}}}$}}

\def\fun#1#2{\lower3.6pt\vbox{\baselineskip0pt\lineskip.9pt\ialign
{$\mathsurround=0pt#1\hfil##\hfil$\crcr#2\crcr\sim\crcr}}}

\newcommand{\lsim}{\ \raise -2.truept\hbox{\rlap{\hbox{$\sim$}}
\raise5.truept \hbox{$<$}\ }}
\newcommand{\gsim}{\ \raise -2.truept\hbox{\rlap{\hbox{$\sim$}}
\raise5.truept\hbox{$>$}\ }}

\begin{document}
\input feynman
\begin{titlepage}
\hskip 12cm \vbox{\hbox{BUDKERINP/96-52}\hbox{CS-TH 4/96}\hbox{July 1996}}
\vskip 0.3cm
\centerline{\bf GRIBOV'S THEOREM ON SOFT EMISSION}
\centerline{\bf AND THE REGGEON-REGGEON-GLUON VERTEX}
\centerline{\bf AT SMALL TRANSVERSE MOMENTUM$^{~\ast}$}
\vskip 0.8cm
\centerline{  V.S. Fadin$^{\dagger}$}
\vskip .1cm
\centerline{\sl Budker Institute for Nuclear Physics}
\centerline{\sl and Novosibirsk State University, 630090 Novosibirsk,
Russia}
\vskip .4cm
\centerline{  R. Fiore$^{\ddagger}$}
\vskip .1cm
\centerline{\sl  Dipartimento di Fisica, Universit\`a della Calabria}
\centerline{\sl Istituto Nazionale di Fisica Nucleare, Gruppo collegato di
Cosenza}
\centerline{\sl Arcavacata di Rende, I-87030 Cosenza, Italy}
\vskip .4cm
\centerline{  M.I. Kotsky$^{\dagger}$}
\vskip .1cm
\centerline{\sl Budker Institute for Nuclear Physics}
\centerline{\sl  630090 Novosibirsk, Russia}
\vskip 0.8cm
\begin{abstract}
In spite of the fact that the Gribov's theorem about the region of 
applicability of the soft emission factorization cannot be referred 
literally to the case of massless charged particles, it can be used for the 
calculation of soft emission amplitudes in such a case also. We demonstrate 
this for the gluon production amplitude in the multi-Regge kinematics with 
small transverse momentum.
\end{abstract}
\vskip .5cm
\hrule
\vskip.3cm
\noindent

\noindent
$^{\ast}${\it Work supported in part by the Ministero italiano 
dell'Universit\`a e della Ricerca Scientifica e Tecnologica, in part by the
EEC Programme ``Human Capital and Mobility", Network ``Physics at High 
Energy Colliders", contract CHRX-CT93-0357 (DG 12 COMA), in part by INTAS 
and in part by the Russian Fund of Basic Researches.}
\vfill
\vskip .2cm
$ \begin{array}{ll}
^{\dagger}\mbox{{\it email address:}} &
 \mbox{FADIN, KOTSKY~@INP.NSK.SU}\\
\end{array}
$

$ \begin{array}{ll}
^{\ddagger}\mbox{{\it email address:}} &
  \mbox{FIORE~@FIS.UNICAL.IT}
\end{array}
$
\vfill
\end{titlepage}
\eject
\textheight 220mm
\baselineskip=24pt

In 1967 V.N. Gribov showed \cite{GB} that at high energy hadron collisions 
the region of applicability of well-known formulas for accompanying 
bremstrahlung extends considerably. Namely, for  collision of two 
particles, A and B, with large c.m.s. energy $\sqrt{s}=\sqrt{(p_A+p_B)^2}$ 
this region is restricted by the inequalities
\beqn
\frac{2p_Ak}{s} \ll 1~,~~~~~~~~~~~~\frac{2p_Bk}{s} \ll 1~,
\zenn
\beq
\vec k_{\perp}^{~2} \approx \frac{2(p_Ak)\cdot 2(p_Bk)}{s} \ll \mu^2~,
\label{z1}
\zen
where $\vec k_\perp$ is the projection of the momentum of the emitted photon 
on the plane orthogonal to the momenta of the colliding particles $p_A$ and 
$p_B$, and $\mu$ is a typical hadron mass.
\vskip.03cm
Gribov proved \cite{GB} that in the region (\ref{z1}) the amplitude of the 
emission process is given only by those Feynman diagrams where the photon line 
is attached to external charged particles. Furthermore, calculating the 
contributions of these diagrams one has to keep the non radiative part of 
the amplitude on the mass shell, i.e. to neglect virtualities of radiating
particles. In the following we shall call the formulas obtained in such a way 
``soft insertion formulas". Let us stress that these formulas are invariant 
under gauge transformation of the emitted photon.
\vskip.03cm
Notice that before the work of Ref. \cite{GB} it was generally accepted 
(see, for example, Ref. \cite{LOW})
that for the applicability of the soft insertion formulas one has to have 
\beq
2p_Ak \ll \mu^2~,~~~~~~~~~~~~2p_Bk \ll \mu^2~.
\label{z2}
\zen
Indeed, the conditions (\ref{z2}) are much more stringent than the 
conditions (\ref{z1}). The possibility of using the factorized formulas with 
the on mass-shell non radiative amplitude in the region (\ref{z1}) is quite 
non trivial and is connected with gauge invariance of the emission amplitude 
\cite{GB}. 
\vskip.03cm
It is very attractive to make use of the Gribov's theorem in more 
complicated cases, such as, for example, Quantum Chromodynamics (QCD). An 
evident obstacle for this is the masslessness of particles having colour 
charge. In other words, the typical mass $\mu$ in Eq. (\ref{z1}) is equal to 
zero for the case of QCD.
\vskip.03cm
The main point in the proof of the Gribov's theorem is the smallness of the 
transverse momentum $k_\perp$ of the emitted quantum of the gauge field 
(photon or gluon) in comparison with the essential transverse momenta of the 
other particles. In massive theories the latter momenta are of order (or 
larger than) $\mu$. Contrary, in theories with massless particles, such as 
QCD, the essential transverse momenta of virtual particles can be arbitrary 
small (that appears as infrared and  collinear divergences). Therefore, the 
Gribov's theorem cannot be applied literally for these theories.
\vskip 0.3cm
Nevertheless, the theorem can be used for the calculation of emission 
amplitudes. Below we demonstrate this for the process of emission of a gluon 
$G$ with momentum $p_G \equiv k$ at scattering of the  particles (quarks or 
gluons) $A$ and $B$,
\beq
A + B \rightarrow A' +B' +G~,
\label{z3}
\zen
in the multi-Regge kinematics
\beqn
s = (p_A + p_B)^2 \gg s_{1,2} \gg |t_{1,2}|~,
\zenn
\beqn
s_1 = (2p_{A'}+k)^2\approx 2p_Ak~,~~~~~~~~~~s_2 = (2p_{B'}+k)^2\approx 2p_Bk~,
\zenn
\beq
t_i = q_i^2 \approx -\vec q_{i\perp}^{~2}~,~~~~~~~~q_1 =p_A - p_{A'}~,~~~~~~~~
q_2 = p_{B'} - p_{B}~,
\label{z4}
\zen
for the case of the transverse momentum of the emitted gluon small compared 
with the transferred momenta:
\beqn
|\vec k_{\perp}| \ll |\vec q_{\perp}|~,~~~~~~~~~~~~q \equiv \frac{q_1+q_2}{2}~, 
\zenn
\beq
|t_1-t_2| \ll |t| \approx \vec q_{\perp}^{~2}~.
\label{z5}
\zen
\vskip 0.3cm
This process is chosen because of its connection with the small $x$ behaviour 
of parton distributions ($x$ is the fraction of the hadron momentum carried by 
a parton). In the leading $ln(1/x)$ approximation (LLA) these distributions 
can be calculated using the BFKL equation \cite{FKL}.
To define a region of its applicability as well as to fix a scale of 
virtualities of the running coupling constant $\alpha_s(Q^2)$ in the equation 
one has to know radiative corrections to the kernel of the equation. The 
program of calculation of the next-to-leading corrections was developed 
in Ref. \cite{LF}. It is based on the gluon Reggeization in QCD; therefore, 
the corrections to the kernel include the one-loop corrections to the 
Reggeon-Reggeon-gluon (RRG) vertex, which are determined by the gluon 
production amplitude in the multi-Regge kinematics. The corrections to 
the RRG vertex were calculated in Refs. [5,6]. The calculations were performed 
in the space-time dimension $D\neq 4$ for regularizing the infrared and 
collinear divergences, but terms vanishing at $D\rightarrow 4$ were omitted in 
the final expressions. Unfortunately, such terms can give non vanishing 
contributions to the total cross sections (and to corrections to the kernel 
of the BFKL equation) because the integration over the transverse momenta of 
the produced gluon leads to divergences at $k_\perp=0$ for the case $D=4$. 
Therefore, in the region $k_\perp \rightarrow0$ we need to know the production 
amplitude for arbitrary $D\neq 4$. As we shall see, the use of the Gribov's 
theorem simplifies considerably the calculation of the amplitude in this region.
\vskip 0.3cm
Let us start with the Born approximation. Obviously, the soft insertion 
formula should be valid here in the region defined by Eqs. (\ref{z4}) and 
(\ref{z5}), because all transverse momenta are fixed and $k_\perp$ is the 
smallest one. The elastic scattering amplitude in the region of large $s$ and 
fixed $t$ in the Born approximation has the form 
\beq
{\cal A}^{A'B'}_{AB}(Born) = 
\Gamma^{(0)i}_{A'A}\frac{2s}{t}\Gamma^{(0)i}_{B'B}~,
\label{z6}
\zen
where $t=-\vec q_\perp^{~2}$ and $\Gamma^{(0)i}_{A'A}$ are the 
particle-particle-Reggeon (PPR) vertices  in the Born approximation 
\cite{FKL}. In the helicity basis these vertices can be presented as 
\beq
\Gamma^{(0)i}_{A'A} = g\langle A'| T^i|A\rangle 
\delta_{\lambda_{A'},\lambda_A}~,
\label{z7}
\zen
where $\langle A'| T^i|A\rangle$ are the matrix elements of the colour group 
generators in the corresponding representation. It is easy to see that the soft 
insertion of a gluon with momentum $p_G\equiv k$, colour index $c$ and 
polarization vector $e(k)$ gives us
\beq
{\cal A}^{A'GB'}_{AB}(Born) = 
\Gamma^{(0)i_1}_{A'A}\frac{2s}{t}\Gamma^{(0)i_2}_{B'B}gT^c_{i_2i_1}e^*_
{\mu}(k)\left(\frac{p_A^\mu}{p_Ak} - \frac{p_B^\mu}{p_Bk}\right)~.
\label{z8}
\zen
Let us remind that in the kinematics defined by the relations (\ref{z4}) the 
gluon production amplitude in the Born approximation takes the form 
\cite{FKL}
\beq
{\cal A}^{A'GB'}_{AB}(Born) = 
2s\Gamma^{(0)i_1}_{A'A}\frac{1}{t_1}gT^c_{i_2i_1}e^*_{\mu}(k)C^\mu(q_2,q_1)
\frac{1}{t_2}\Gamma^{(0)i}_{B'B}~,
\label{z9}
\zen
where the effective production vertex is
\beq
C(q_2,q_1) = -q_{1\perp}-q_{2\perp}+ p_A\left(\frac{q^2_1}{p_Ak} + 
\frac{p_Bk}{p_Ap_B}\right) - p_B\left(\frac{q^2_2}{p_Bk} + 
\frac{p_Ak}{p_Ap_B}\right)~. 
\label{z10}
\zen
In the region (\ref{z5}) of small $k_\perp$ we find that
\beq
C(q_2,q_1) \rightarrow t\left(\frac{p_A}{p_Ak} - \frac{p_B}{p_Bk}\right)~,
\label{z11}
\zen
and the expression (\ref{z9}) turns into the form (\ref{z8}). So, for the case 
of the Born approximation the soft insertion formula is valid, as it was 
reported.
\vskip 0.3cm
Now let us consider the one-loop corrections to the production amplitude. 
Since in this case we need to integrate over the momenta of virtual particles, 
we cannot expect that the soft insertion gives a corrected answer here. 
However, analyzing the proof of the Gribov's theorem \cite{GB} one can 
conclude that the soft insertion should be  valid  for the contribution of the 
kinematical region where the transverse momenta of virtual particles are much 
larger than $k_\perp$. The idea is to use the soft insertion formula for this 
contribution and to add the contribution of the region of small virtual 
transverse momenta, which has to be calculated separately. From the first 
sight the idea appears doubtful, because for $D=4$ the integrals over virtual 
transverse momenta have a logarithmic behaviour; therefore, it seems that the 
separation of two regions is not a simple problem. But for $D>4$ the integrals 
are convergent, and we have two different scales where they can converge, 
$q_\perp$ and $k_\perp$, so that the separation is quite simple in this 
case. Evidently, the contribution of the integrals converging at $q_\perp$ can 
be obtained applying the soft insertion formula and the contribution 
of the integrals converging at $k_\perp$ has to be calculated.
\vskip 0.3cm
Fortunately, a simple inspection of the Feynman diagrams shows that only those 
ones of Fig. 1 lead to the integrals of the second kind. This statement is 
valid for all possible choices of colliding particles: they can be gluons (in 
this case all lines in the diagrams of Fig. 1 are gluon lines) or quarks (in 
this case the upper and lower lines  in the diagrams are quark lines) and so on.
\vskip 0.3cm
\begin{picture}(2000,2000)


\drawline\fermion[\E\REG](0,0)[12000]
\put(-1500,0){$p_A$}
\put(13000,0){$p_{A'}$}
\thicklines
\drawarrow[\LDIR\ATTIP](\pmidx,\pmidy)
\drawarrow[\E\ATBASE](\pbackx,\pbacky)
\drawarrow[\E\ATBASE](1000,0)
\thinlines
\drawline\fermion[\E\REG](0,-7500)[12000]
\put(-1500,-7500){$p_B$}
\put(13000,-7500){$p_{B'}$}
\thicklines
\drawarrow[\LDIR\ATTIP](\pmidx,\pmidy)
\drawarrow[\E\ATBASE](\pbackx,\pbacky)
\drawarrow[\E\ATBASE](1000,-7500)
\thinlines
\drawline\gluon[\S\REG](2000,0)[7]
\drawline\gluon[\S\REG](9000,0)[7]
\drawline\gluon[\E\CURLY](9000,-3750)[3]
\put(12500,-4000){$k$}
\thicklines
\drawarrow[\E\ATBASE](\pbackx,\pbacky)
\put(6000,-9500){$a)$}


\drawline\fermion[\E\REG](24000,0)[12000]
\put(22500,0){$p_A$}
\put(37000,0){$p_{A'}$}
\thicklines
\drawarrow[\LDIR\ATTIP](\pmidx,\pmidy)
\drawarrow[\E\ATBASE](\pbackx,\pbacky)
\drawarrow[\E\ATBASE](25000,0)
\thinlines
\drawline\fermion[\E\REG](24000,-7500)[12000]
\put(22500,-7500){$p_B$}
\put(37000,-7500){$p_{B'}$}
\thicklines
\drawarrow[\LDIR\ATTIP](\pmidx,\pmidy)
\drawarrow[\E\ATBASE](\pbackx,\pbacky)
\drawarrow[\E\ATBASE](25000,-7500)
\thinlines
\drawline\gluon[\S\REG](26000,0)[7]
\drawline\gluon[\S\REG](33000,0)[7]
\drawline\gluon[\E\CURLY](26000,-3750)[3]
\put(29500,-4000){$k$}
\thicklines
\drawarrow[\E\ATBASE](\pbackx,\pbacky)
\put(30000,-9500){$b)$}


\drawline\fermion[\E\REG](0,-13000)[12000]
\put(-1500,-13000){$p_A$}
\put(13000,-13000){$p_{A'}$}
\thicklines
\drawarrow[\LDIR\ATTIP](\pmidx,\pmidy)
\drawarrow[\E\ATBASE](\pbackx,\pbacky)
\drawarrow[\E\ATBASE](1000,-13000)
\thinlines
\drawline\fermion[\E\REG](0,-20500)[12000]
\put(-1500,-20500){$p_B$}
\put(13000,-20500){$p_{B'}$}
\thicklines
\drawarrow[\LDIR\ATTIP](\pmidx,\pmidy)
\drawarrow[\E\ATBASE](\pbackx,\pbacky)
\drawarrow[\E\ATBASE](1000,-20500)
\thinlines
\drawline\gluon[\SE\REG](2000,-13000)[7]
\drawline\gluon[\SW\REG](9000,-13000)[7]
\drawline\gluon[\E\CURLY](7000,-15500)[3]
\put(10750,-15750){$k$}
\thicklines
\drawarrow[\E\ATBASE](\pbackx,\pbacky)
\put(6000,-22500){$c)$}


\drawline\fermion[\E\REG](24000,-13000)[12000]
\put(22500,-13000){$p_A$}
\put(37000,-13000){$p_{A'}$}
\thicklines
\drawarrow[\LDIR\ATTIP](\pmidx,\pmidy)
\drawarrow[\E\ATBASE](\pbackx,\pbacky)
\drawarrow[\E\ATBASE](25000,-13000)
\thinlines
\drawline\fermion[\E\REG](24000,-20500)[12000]
\put(22500,-20500){$p_B$}
\put(37000,-20500){$p_{B'}$}
\thicklines
\drawarrow[\LDIR\ATTIP](\pmidx,\pmidy)
\drawarrow[\E\ATBASE](\pbackx,\pbacky)
\drawarrow[\E\ATBASE](25000,-20500)
\thinlines
\drawline\gluon[\SE\REG](26000,-13000)[7]
\drawline\gluon[\SW\REG](33000,-13000)[7]
\drawline\gluon[\E\FLIPPED](31500,-18000)[3]
\put(35250,-18250){$k$}
\thicklines
\drawarrow[\E\ATBASE](\pbackx,\pbacky)
\put(30000,-22500){$d)$}

\put(16000,-25000){$Fig.~~1$}

\end{picture}
\vskip 10.0cm
\begin{description}
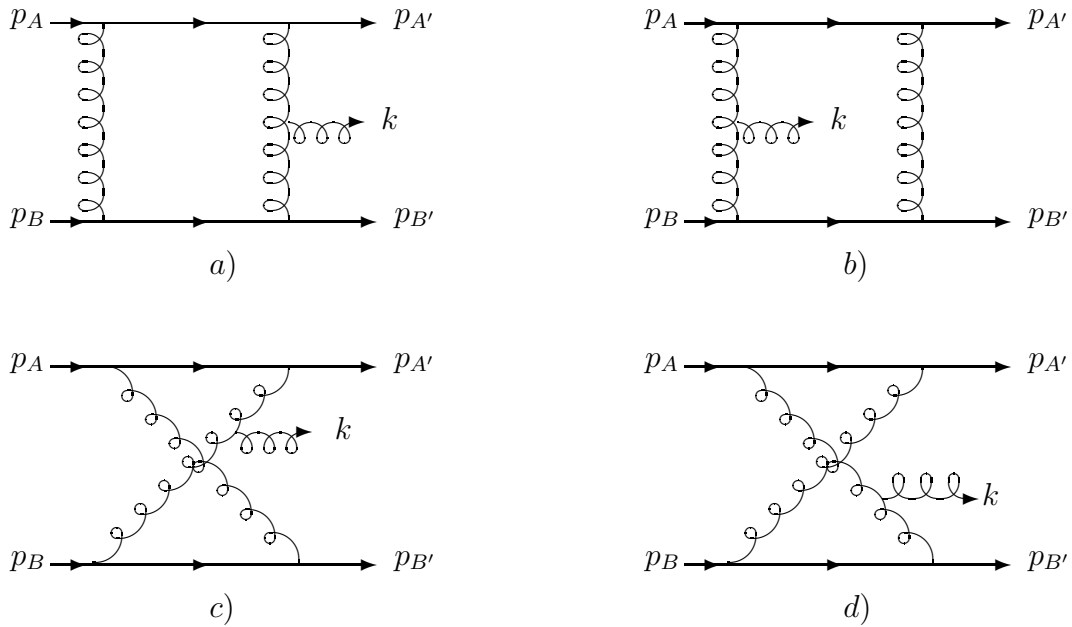
\item{Fig. 1:}
Feynman diagrams giving a non factorizable contribution to the gluon
emission amplitude.
\end{description}

Let us split the production amplitude as the sum of the factorizable and 
non contributions:
\beq
{\cal A}^{A'GB'}_{AB} = {\cal A}^{A'GB'}_{AB}(f) + {\cal A}^{A'GB'}_{AB}(nf)~.
\label{z12}
\zen
The first term in Eq. (\ref{z12}) comes from the soft insertion while the 
second one is represented in the one-loop approximation by the diagrams of 
Fig. 1.
\vskip 0.3cm
Contrary to the Born case, in higher orders the colour structure of the 
production amplitude is not so simple. For definiteness, let us consider the 
part of the amplitude with the gluon quantum numbers in $t_1$ and $t_2$ 
channels. This part is the most important one because it determines the RRG 
vertex. The factorizable contribution to this part has a form similar to the 
expression (\ref{z8}):
\beq
{\cal A}^{(8)~A'GB'}_{AB}(f) = 
\Gamma^{i_1}_{A'A}\left[\left(\frac{-s}{-t}\right)^{j(t)} - \left(\frac{s}{-t}
\right)^{j(t)}\right]\Gamma^{i_2}_{B'B}gT^c_{i_2i_1}e^*_{\mu}(k)
\left(\frac{p_A^\mu}{p_Ak} - \frac{p_B^\mu}{p_Bk}\right)~.
\label{z13}
\zen
Here $j(t)=1+\omega(t)$ is the gluon trajectory \cite{FKL} and 
$\Gamma^i_{A'A}$ are the PPR vertices. The one-loop corrections to the LLA 
vertices (\ref{z7}) are calculated in Refs. [5,7]. There one can find an 
explicit expression for $\omega(t)$ also.
\vskip 0.3cm
Now let us pass to the non factorizable contribution. Evidently, the 
diagrams of Fig. 1 are connected each other by crossing, therefore it is 
sufficient to calculate the contribution of the diagram a). Performing usual 
tricks with the numerators of the gluon propagators connecting lines with 
strongly different momenta:
\beq
g^{\mu\nu} \rightarrow \frac{2p^\mu_Ap^\nu_B}{s}~,
\label{z14}
\zen
and simplifying the numerators of the integrand allow to present the 
contribution of the diagram a) of Fig. 1 in the form 
\beq
{\cal A}^{(8)~A'GB'}_{AB}(a) = 
-\frac{g^3N}{4}\Gamma^{(0)~i_1}_{A'A}\Gamma^{(0)~i_2}_{B'B}T^c_{i_2i_1}e^*_
{\mu}(k)\left(\frac{p_A^\mu}{p_Ak}-\frac{p_B^\mu}{p_Bk}\right)ss_1s_2\cal I~,
\label{z15}
\zen
where 
\beq
{\cal I} = \int \frac{d^Dp}{(2\pi)^Di(p^2+i\varepsilon)((p+p_A)^2+i\varepsilon)
((p-p_B)^2+i\varepsilon)((p+q_1)^2+i\varepsilon)((p+q_2)^2+i\varepsilon)}~.
\label{z16}
\zen
In the region defined by the relations (\ref{z4}) and (\ref{z5}) we get
\beq
{\cal I} = 
-\frac{1}{(4\pi)^{\frac{D}{2}}}\frac{\Gamma^2\left(3-\frac{D}{2}\right)
\Gamma^3\left(\frac{D}{2}-2\right)}{s_1s_2\vec q^{~2}\Gamma(D-4)}
\left(-\frac{s_1s_2}{s}\right)^{\frac{D}{2}-2}~.
\label{z17}
\zen
Consequently, using a simple colour algebra and the crossing relations we 
obtain
\beqn
 {\cal A}^{(8)~A'GB'}_{AB}(nf) = 
\Gamma^{(0)~i_1}_{A'A}\frac{2s}{t}\Gamma^{(0)~i_2}_
{B'B}gT^c_{i_2i_1}e^*_{\mu}(k)\left(\frac{p_A^\mu}{p_Ak}-
\frac{p_B^\mu}{p_Bk}\right)(-\frac{g^2N}{8(4\pi)^{\frac{D}{2}}})
(\vec k_\perp^{~2})^{\frac{D}{2}-2}
\zenn
\beq
\times \left[3exp\left(-i\pi\left(\frac{D}{2}-
2\right)\right) + exp\left(i\pi\left(\frac{D}{2}-2\right)\right)\right]
\frac{\Gamma^2\left(3-\frac{D}{2}\right)
\Gamma^3\left(\frac{D}{2}-2\right)}{\Gamma(D-4)}~.
\label{z18}
\zen

The total amplitude is given by the sum of Eqs.(13) and (18).
\vskip 0.3cm
Assuming the Regge behaviour of the amplitude in the sub-channels $s_1$ and 
$s_2$, from general requirements of analiticity, unitarity and crossing 
symmetry one has (see Refs. \cite{BLF,BAR})
\beqn
{\cal A}^{(8)~A'GB'}_{AB} = 
s\Gamma^{i_1}_{A'A}\frac{1}{t_1}T^c_{i_2i_1}\frac{1}{t_2}\Gamma^{i_2}_{B'B}
\zenn
\beqn
\times \left\{\frac{1}{4}\left[
\left(\frac{-s_1}{\mu^2}\right)^{\omega_1-\omega_2} + 
\left(\frac{s_1}{\mu^2}\right)^{\omega_1-\omega_2}
\right]\left[
\left(\frac{-s}{\mu^2}\right)^{\omega_2} +
\left(\frac{s}{\mu^2}\right)^{\omega_2}
\right]R\right.
\zenn
\beq
\left.+\frac{1}{4}\left[
\left(\frac{-s_2}{\mu^2}\right)^{\omega_2-\omega_1} +
\left(\frac{s_2}{\mu^2}\right)^{\omega_2-\omega_1}
\right]\left[
\left(\frac{-s}{\mu^2}\right)^{\omega_1} +
\left(\frac{s}{\mu^2}\right)^{\omega_1}
\right]L\right\}~,
\label{z19}
\zen
where $\omega_i=\omega(t_i)$ and the RRG vertices 
R and L are real in all physical channels. 
In the region (5) of small $k_{\perp}$ this representation reduces to 
\beqn
{\cal A}^{(8)~A'GB'}_{AB} = 
s\Gamma^{i_1}_{A'A}\frac{1}{t}T^c_{i_2i_1}\frac{1}{t}\Gamma^{i_2}_{B'B}
\left\{\left[
\left(\frac{-s}{\mu^2}\right)^{\omega} +
\left(\frac{s}{\mu^2}\right)^{\omega}
\right]\frac{R+L}{2}\right.
\zenn
\beq
\left.+\left[
\left(\frac{s}{\mu^2}\right)^{\omega}\left(\ln\left(\frac{\vec k_{\perp}^{~2}}
{\mu^2}\right)-i\pi\right) +
\left(\frac{-s}{\mu^2}\right)^{\omega}
\ln\left(\frac{\vec k_{\perp}^{~2}}
{\mu^2}\right)\right]\frac{(\omega_1-\omega_2)}{2}\frac{(R-L)}{2}\right\}.
\label{z20}
\zen
Comparing the above form with Eqs. (13) and (18) we conclude that  
\beqn
 R-L = 
ge^*_{\mu}(k)\left(\frac{p_A^\mu}{p_Ak}-
\frac{p_B^\mu}{p_Bk}\right)\frac{t}{\omega_1-\omega_2}
\zenn
\beqn
\times \left(-\frac{2g^2N}{(4\pi)^{\frac{D}{2}}}\right)
(\vec k_{\perp}^{~2})^{\frac{D}{2}-2}
\frac{\Gamma^2\left(3-\frac{D}{2}\right)
\Gamma^3\left(\frac{D}{2}-2\right)}{\Gamma(D-4)}
\frac{\sin\left(\pi\left(\frac{D}{2}-
2\right)\right)}{\pi}~,
\zenn
\beqn
 R+L = 
2ge^*_{\mu}(k)\left(\frac{p_A^\mu}{p_Ak}-
\frac{p_B^\mu}{p_Bk}\right)t\left\{1-\omega(t)\ln\left(\frac{-t}{\mu^2}\right)
-\frac{g^2N}{(4\pi)^{\frac{D}{2}}}
\frac{\Gamma^2\left(3-\frac{D}{2}\right)
\Gamma^3\left(\frac{D}{2}-2\right)}{2\Gamma(D-4)}
\right.
\zenn
\beq
\left. \times 
(\vec k_{\perp}^{~2})^{\frac{D}{2}-2}
\left[\cos\left(\pi\left(\frac{D}{2}-
2\right)\right)-\frac{\sin\left(\pi\left(\frac{D}{2}-
2\right)\right)}{\pi}\ln\left(\frac{\vec k_{\perp}^{~2}}{\mu^2}\right)
\right]\right\}~.
\label{z21}
\zen
At $D\rightarrow 4$ the above expressions for the vertices coincide,  
taking into account  the charge renormalization, with the small $k_{\perp}$ 
limit of the corresponding expressions of Ref. \cite {BLF} (see Eq.(86) there). 
Independently we have performed the straightforward calculation of the RRG 
vertices at small 
$k_{\perp}$ for arbitrary $D$ and have obtained the result (21).
\vskip 0.3cm
In conclusion we stress  that although the Gribov's theorem about the soft 
emission factorization cannot be literally applied in the case of massless 
theories, it can simplify calculations considerably.

\vskip 1.5cm
\underline {Acknowledgement}: One of us (V.S.F.) thanks the Dipartimento di
Fisica della Universit\`a della Calabria and the Istituto Nazionale di 
Fisica Nucleare - Gruppo collegato di Cosenza for their warm hospitality 
while part of this work was done.

\end{document}